# Revisiting the Complexity of And/Or Graph Solution

Maise Dantas da Silva, Fábio Protti, Uéverton dos Santos Souza

*Fluminense Federal University, Niterói, RJ, Brazil*

maisedantas@id.uff.br, fabio@ic.uff.br, usouza@ic.uff.br

**Abstract**

This paper presents a study on two data structures that have been used to model several problems in computer science: and/or graphs and x-y graphs. An and/or graph is an acyclic digraph containing a source (a vertex that reaches all other vertices by directed paths), such that every vertex $v$ has a label $f(v) \in \{\texttt{and},\texttt{or}\}$ and (weighted) edges represent dependency relations between vertices: a vertex labeled $\texttt{and}$ depends on all of its out-neighbors (conjunctive dependency), while a vertex labeled $\texttt{or}$ depends on only one of its out-neighbors (disjunctive dependency). X-y graphs are defined as a natural generalization of and/or graphs: every vertex $v_i$ of an x-y graph has a label $x_i$-$y_i$ to mean that $v_i$ depends on $x_i$ of its $y_i$ out-neighbors. We analyze the complexity of the optimization problems MIN-AND/OR and MIN-X-Y, which consist of finding solution subgraphs of optimal weight for and/or and x-y graphs, respectively. A solution subgraph $H$ of an and/or-graph must contain the source and obey the following rule: if an $\texttt{and}$-vertex (resp. $\texttt{or}$-vertex) is included in $H$ then all (resp. one) of its out-edges must also be included in $H$. Analogously, if a vertex $v_i$ is included in a solution subgraph $H$ of an x-y graph then $x_i$ of its $y_i$ out-edges must also be included in $H$. Motivated by the large applicability as well as the hardness of MIN-AND/OR and MIN-X-Y, we study new complexity aspects of such problems, both from a classical and a parameterized point of view. We prove that MIN-AND/OR remains NP-hard even for a very restricted family of and/or graphs where edges have weight one and $\texttt{or}$-vertices have out-degree at most two (apart from other property related to some in-degrees), and that deciding whether there is a solution subtree with weight exactly $k$ of a given x-y tree is also NP-hard. We also show that: (i) the parameterized problem MIN-AND/OR($k,r$), which asks whether there is a solution subgraph of weight at most $k$ where every $\texttt{or}$-vertex has at most $r$ out-edges with the same weight, is FPT; (ii) the parameterized problem MIN-AND/OR$^0(k)$, whose domain includes and/or graphs allowing zero-weight edges, is W[2]-hard; (iii) the parameterized problem MIN-X-Y($k$) is W[1]-hard.

## 1 Introduction

In this paper we consider the complexity of problems involving two important data structures, *and/or graphs* and *x-y graphs*. An and/or graph is an acyclic digraph containing a source (a vertex that reaches all other vertices by directed paths), such that every vertex $v \in V(G)$ has a label $f(v) \in \{\texttt{and},\texttt{or}\}$. In such digraphs, edges represent dependency relations between vertices: a vertex labeled `and` depends on all of its out-neighbors (conjunctive dependency), while a vertex labeled `or` depends on only one of its out-neighbors (disjunctive dependency).

We define x-y graphs as a generalization of and/or graphs: every vertex $v_i$ of an x-y graph has a label $x_i$-$y_i$ to mean that $v_i$ depends on $x_i$ of its $y_i$ out-neighbors. Given an and/or graph $G$, an equivalent x-y graph $G'$ is easily constructed as follows: sinks of $G$ are vertices with $x_i = y_i = 0$; `and`-vertices satisfy $x_i = y_i$; and `or`-vertices satisfy $x_i = 1$.

In representations of and/or graphs, `and`-vertices have an arc around its out-edges. Figure 1 shows in (a) an example of and/or graph, and in (b) an example of x-y graph.

And/or graphs were used for modeling problems originated in the 60's within the domain of Artificial Intelligence [17, 19]. Since then, they have successfully been applied to other fields, such as Operations Research, Automation, Robotics, Game Theory, and Software Engineering, to model cutting problems [15], interference tests [11], failure dependencies [4], robotic task plans [5], assembly/disassembly sequences [7], game trees [13], software versioning [6], and evaluation of boolean formulas [14]. With respect to x-y graphs, they correspond to the *x-out-of-y model* of resource sharing in distributed systems [3].

In addition to the above applications, special directed hypergraphs named *F-graphs* are equivalent to and/or graphs [10]. An F-graph is a directed hypergraph where hyperarcs are called *F-arcs* (for *forward arcs*), which are of the form $E_i = (S_i, T_i)$ with $|S_i| = 1$. An F-graph $H$ can be easily transformed into an and/or graph as follows: for each vertex $v \in V(H)$ do $f(v)=\texttt{or}$; for each $F$-arc $E_i = (S_i, T_i)$, where $|T_i| \geq 2$, do: create an `and`-vertex $v_i$, add an edge $(u, v_i)$ where $\{u\} = S_i$, and add an edge $(v_i, w_j)$ for all $w_j \in T_i$.

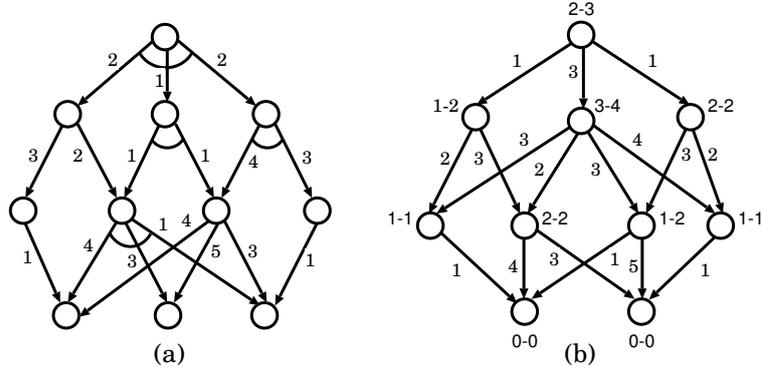

Figure 1: (a) A weighted and/or graph; (b) A weighted x-y graph.

In this work, we denote by $O_v$ and $I_v$, respectively, the subsets of out-neighbors and in-neighbors of a vertex $v$. Also, $\tau(e)$ denotes the weight of an edge $e$, and we define the weight of a graph as the sum of the weights of its edges. We assume $|V(G)| = n$ and $|E(G)| = m$.

The optimization problems associated with and/or graphs and x-y graphs are formally defined below.

Min-and/or
*Instance:* An and/or graph $G = (V, E)$ where each edge $e$ has an integer weight $\tau(e) > 0$.
*Goal:* Determine the minimum weight of a subdigraph $H = (V', E')$ of $G$ (*solution subgraph*) satisfying the following properties:
• $s \in V'$;
• if a non-sink node $v$ is in $V'$ and $f(v)=$and then every out-edge of $v$ belongs to $E'$;
• if a non-sink node $v$ is in $V'$ and $f(v)=$or then exactly one out-edge of $v$ belongs to $E'$.

Min-x-y
*Instance:* An x-y graph $G = (V, E)$ where each edge $e$ has an integer weight $\tau(e) > 0$.
*Goal:* Determine the minimum weight of a subdigraph $H = (V', E')$ of $G$ satisfying the following properties:
• $s \in V'$;
• for every non-sink node $v_i$ in $V'$, $x_i$ of its $y_i$ out-edges belong to $E'$.

In 1974, Sahni [18] showed that Min-and/or is NP-hard via a reduction from 3-SAT. Therefore, MIN-X-Y is also NP-hard.

There are three trivial cases for which MIN-AND/OR can be solved in polynomial time:

1. All vertices of $G$ are and-vertices. In this case, $G$ is the solution subgraph.

2. All vertices of $G$ are or-vertices. In this case, the optimal solution subgraph is a shortest path between $s$ and a sink.

3. $G$ is a tree (*and/or tree*). In this case, the weight of the optimal solution subgraph of $G$, given by $c(s)$, can be obtained in $O(n)$ time via the recurrence relation below:

$$c(v_i) = \begin{cases} 0, \text{if } v_i \text{ is a sink;} \\ \sum_{v_j \in O_{v_i}} (\tau(v_i, v_j) + c(v_j)), \text{ if } f(v_i) = \text{and;} \\ \min_{v_j \in O_{v_i}} \{\tau(v_i, v_j) + c(v_j)\}, \text{ if } f(v_i) = \text{or.} \end{cases}$$

Other three trivial cases of MIN-AND/OR can be listed: if every or-vertex has out-degree one then or-vertices can be converted into and-vertices, and case 1 above applies; if every and-vertex has out-degree one then and-vertices can be converted into or-vertices, and case 2 applies; finally, if every vertex with in-degree greater than 1 is a sink then the recurrence presented in the case 3 can be used.

As noted by Adelson-Velsky in [1], the problem MIN-AND/OR has interesting connections with real-word applications in scheduling. An example is the work [2], which employs and/or graphs to model real-time scheduling of tasks in computer communication systems. Such a scheduling problem (AND/OR-SCHEDULING) generalizes the classical shortest-path and critical-path problems in graphs [1]. Given a weighted and/or graph, AND/OR-SCHEDULING consists of finding the earliest starting times $t(v_i)$, for all $v_i \in V(G)$, satisfying the following conditions:

- $t(v_i) = 0$, if $v_i$ is a sink;

- $t(v_i) \geq \max_{v_j \in O_{v_i}} \{\tau(v_i, v_j) + t(v_j)\}$, if $f(v_i) = \text{and}$;

- $t(v_i) \geq \min_{v_j \in O_{v_i}} \{\tau(v_i, v_j) + t(v_j)\}$, if $f(v_i) = \text{or}$.

MIN-AND/OR can thus be viewed as a variant of AND/OR-SCHEDULING: while the latter aims at determining the minimum *time* necessary to perform a task, the former aims at determining the minimum *cost* to perform it. Since AND/OR-SCHEDULING is solvable in polynomial time [1], its solution can be used as a practical lower bound for MIN-AND/OR. In addition, the recurrence equations for and/or trees lead to a bottom-up dynamic programming algorithm to find in polynomial time a feasible solution (and hence an upper bound) of MIN-AND/OR.

An *x-y tree* is an x-y graph where no two vertices share a common out-neighbor. As for MIN-AND/OR, MIN-X-Y can be solved in $O(n)$ time when the input x-y graph is an x-y tree $T = (V, E)$. To show this, observe first that the minimum weight of a solution subtree is given by a similar recurrence (shown below), since the optimal solution of an x-y tree rooted at a vertex $v_i$ is obtained by $x_i$ subtrees of $v_i$:

$$c(v_i) = \begin{cases} 0, \text{ if } v_i \text{ is a sink;} \\ \min_{X \subseteq O_{v_i}, |X|=x_i} \left\{ \sum_{x \in X} (\tau(v_i, x) + c(x)) \right\} \end{cases}$$

For each non-sink $v_i$, we need to compute the sum of the $x_i$ smallest values $\tau(v_i, x) + c(x)$ among its children; determining the $x_i$-th smallest value takes $O(y_i)$ time, and thus selecting the $x_i$ smallest values takes $O(y_i)$ time as well. Then the entire bottom-up procedure takes overall $\sum_{i=1}^{n} O(y_i) = O(n)$ time.

Motivated by the large applicability as well as the hardness of MIN-AND/OR and MIN-X-Y, we study new complexity aspects of such problems, both from a classical and a parameterized point of view. The latter is justified by the fact that many applications are concerned with satisfying a low cost limit. The remainder of this work is organized as follows. In Section 2, we prove that MIN-AND/OR remains NP-hard even for a very restricted family of and/or graphs where edges have weight one and or-vertices have out-degree at most two (apart from another property related to some in-degrees), and that deciding whether there is a solution subtree with weight exactly $k$ of a given x-y tree is NP-hard. In Section 3, we show that: (i) the parameterized problem MIN-AND/OR$(k, r)$, which asks whether there is a solution subgraph of weight at most $k$ where every or-vertex has at most $r$ out-edges with the same weight, is FPT; (ii) the parameterized problem MIN-AND/OR$^0(k)$, whose domain includes and/or graphs allowing zero-weight edges, is W[2]-hard; (iii) the parameterized problem MIN-X-Y$(k)$ is W[1]-hard.

## 2  NP-hardness results

We now consider a very restricted family of and/or graphs, defined as follows: Let $\mathcal{F}$ be the set of all and/or graphs $G$ satisfying the following properties: every edge in $E(G)$ has weight one; every or-vertex in $V(G)$ has out-degree at most two; and vertices in $V(G)$ with in-degree greater than one are within distance at most one of a sink. We show that even for such and/or graphs the problem MIN-AND/OR remains NP-hard.

**Theorem 1** MIN-AND/OR *restricted to $\mathcal{F}$ is NP-hard.*

**Proof.**  The proof uses a reduction from VERTEX COVER, shown to be NP-hard by Karp in [12]. Given a graph $G = (V, E)$, we construct an and/or graph $G' = (V', E')$ in $\mathcal{F}$ as follows. Suppose $V = \{v_1, \ldots, v_n\}$ and $E = \{e_1, \ldots, e_m\}$. Create a source $s \in V'$ with $f(s) =$ and. For each edge $e_i \in E$ create an out-neighbor $w_{e_i} \in V'$ of $s$ with $f(w_{e_i}) =$ or. For each vertex $v_j \in V$ create a vertex $w_{v_j} \in V'$ with $f(w_{v_j}) =$ or, and add an edge $(w_{e_i}, w_{v_j})$ in $E'$ if and only if $e_i$ is incident to $v_j$. Finally, create an out-neighbor $t_{v_j}$ for each vertex $w_{v_j} \in V'$ and assign $\tau(e) = 1$ for all $e \in E'$. Figure 2 illustrates in (a) a graph $G$ and in (b) the and/or graph $G'$ obtained by the construction above.

We now show that there is a vertex cover of size at most $k$ in $G$ if and only if there is a solution subgraph of weight at most $2m + k$ in $G'$. Suppose first that $G$ has a vertex cover $C$ of size at most $k$. A suitable solution subgraph $H$ of $G'$ can be obtained as follows. Vertex $s$ must belong to $V(H)$ by definition. Since $s$ is an and-vertex, its $m$ out-edges must belong to $E(H)$. But every out-neighbor $w_{e_i}$ of $s$ is an or-vertex; then exactly one of its out-edges in $G'$, say $(w_{e_i}, w_{v_j})$, must also belong to $E(H)$. We choose edge $(w_{e_i}, w_{v_j})$ if and only if $v_j \in C$. At this point, at most $|C|$ vertices $w_{v_j}$ belong to $V(H)$. Now each $w_{v_j}$ has exactly one out-neighbor which is a sink; then for each $w_{v_j}$ we add only one additional out-edge of it. Hence $H$ has weight $2m + |C| \leq 2m + k$.

Conversely, suppose that $G'$ contains a solution subgraph $H$ of weight at most $2m + k$. By construction, $m$ out-edges of $s$ belong to $E(H)$, and for each vertex $w_{e_i}$ in $V(H)$ exactly one of its out-edges is in $E(H)$. Since each vertex $w_{v_j}$ in $V(H)$ must have one out-neighbor, $V(H)$ contains at most $k$ vertices $w_{v_j}$. Let $X$ be the subset of vertices of the form $w_{v_j}$ in $V(H)$, and $C$ a subset of vertices of $G$ such that $v_j \in C$ if and only if $w_{v_j} \in X$. Every vertex $w_{e_i}$ in $V(H)$ has an out-neighbor $w_{v_j}$ in $V(H)$, and by construction of $G'$ a vertex $w_{e_i}$ is an in-neighbor of $w_{v_j}$ if and only if $e_i$ is incident to $v_j$ in $G$. Since every $w_{e_i}$ in $V(H)$ has an out-neighbor $w_{v_j} \in X$, every edge $e_i$ in $G$ is incident to a vertex $v_j \in C$. Hence $C$ is a vertex cover of $G$ and $|C| = |X| \leq k$. □

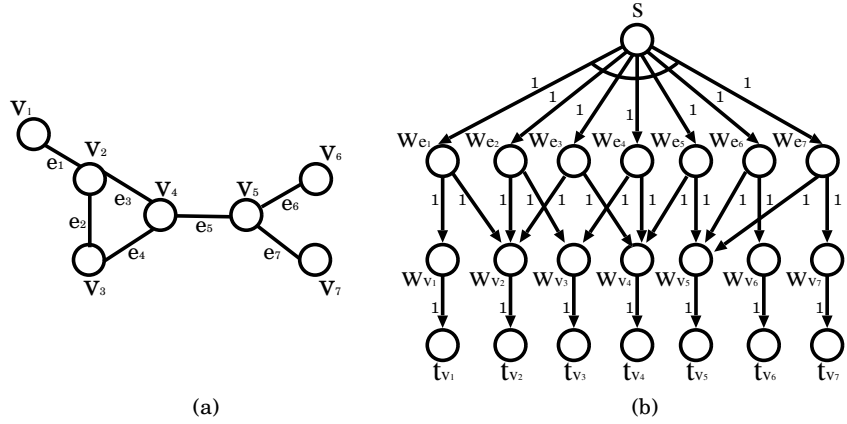

Figure 2: A graph $G$ and the corresponding and/or graph $G'$.

To conclude this section, we show an interesting result concerning x-y trees. Although MIN-X-Y can be solved in linear time when restricted to x-y trees, deciding whether there is a solution subtree with weight exactly $k$ of a given x-y tree is NP-hard.

**Theorem 2** *Let $T$ be an x-y tree. Deciding whether there is a solution subtree $T'$ of $T$ with weight exactly $k$ is NP-hard.*

**Proof.** The proof uses a reduction from the SUBSET SUM problem, shown to be NP-hard by Karp in [12]. It consists of deciding whether in a set of integers there is a subset $S$ of cardinality $p$ such that the sum of the integers in $S$ is equal to an integer value $q$. Given a set of integers $Z = \{z_1, z_2, ..., z_n\}$, an integer $q$ and a positive integer $p$, we construct an x-y tree $T = (V, E)$ such that there is a solution subtree $T'$ of $T$ of weight exactly $k = q + p$ if and only if there is a subset $Z'$ of $Z$ such that $|Z'| = p$ and the sum of the elements in $Z'$ equals $q$. The construction is as follows. Create a source vertex $s \in V(T)$ with label $p$-$n$. For each element $z_i \in Z$, create a vertex $u_i \in V(T)$ with label 1-1 and add an edge $e_i = (s, u_i) \in E(T)$ where $\tau(e_i) = 1$. Finally, for each element $z_i \in Z$, create a vertex $w_i$ with label 0-0 and add an edge $f_i = (u_i, w_i)$ with $\tau(f_i) = z_i$.

Suppose that there is a subset $Z'$ of $Z$ such that $|Z'| = p$ and the sum of its elements equals $q$. Since the source vertex $s$ has label $p$-$n$, a solution subtree $T'$ is constructed as follows: $s \in V(T')$, and for each $z_i \in Z'$ add edges $(s, u_i)$ and $(u_i, w_i)$ to $E(T')$, where $u_i$ and $w_i$ are vertices associated with $z_i$ by construction. Observe that each out-edge $e_i$ of $s$ satisfies $\tau(e_i) = 1$, and each edge $f_i = (u_i, w_i)$ satisfies $\tau(f_i) = z_i$. Hence the weight of $T'$ is $k = q + p$.

Conversely, suppose that there is a solution subtree $T'$ of $T$ with weight $p+q$. By definition, $s \in V(T')$, and there are $p$ out-edges $e_i$ of $s$ belonging to $E(T')$, each one with weight equal to 1. Let $E'$ be the subset of edges of the form $f_i = (u_i, w_i)$ in $E(T')$. Note that $|E'| = p$ and $\sum_{f_i \in E'} \tau(e_i) = q$. Define $Z' = \{z_i \in Z \mid f_i = (u_i, w_i) \in E'\}$. Clearly, $|Z'| = p$ and $\sum_{z_i \in Z'} z_i = q$. □

## 3 Parameterized complexity results

The Parameterized Complexity Theory was proposed by Downey and Fellows [8] as a promising alternative to deal with NP-hard problems described by the following general form [16]: given an object $x$ and a nonnegative integer $k$, does $x$ have some property that depends only on $k$ (and not on the size of $x$)? In parameterized complexity theory, $k$ is fixed as the *parameter*, considered to be *small* in comparison with the size $|x|$ of object $x$. It may be of high interest for some problems to ask whether they admit deterministic algorithms whose running times are exponential with respect to $k$ but polynomial with respect to $|x|$.

**Definition 1** [9] *A parameterized problem $\Pi$ is fixed-parameter tractable, or FPT, if the question "$(x, k) \in \Pi$?" can be decided in running time $f(|k|).|x|^{O(1)}$, where $f$ is an arbitrary function on nonnegative integers. The corresponding complexity class is called FPT.*

**Definition 2** [9] *Let $\Pi = (I, k)$ be a parameterized problem, where instance $I$ is asked to have a solution of size $k$. Reduction to problem kernel means to replace instance $(I, k)$ by a reduced instance $(I', k')$ (called problem kernel) such that $k' \leq ck$ for a constant $c$, $|I'| \leq g(k)$ for some function $g$ only depending on $k$, and $(I, k) \in \Pi$ if and only if $(I', k') \in \Pi$. Furthermore, the reduction from $(I, k)$ to $(I', k')$ is computable in polynomial time.*

**Definition 3** [9] *Let $(Q, k)$ and $(Q', k')$ be parameterized problems over alphabets $\Sigma$ and $\Sigma'$, respectively. An FPT-reduction from $(Q, k)$ to $(Q', k')$ is a mapping $R : \Sigma^* \to (\Sigma')^*$ such that:*

1. *For all $x \in \Sigma^*$, it holds that $x \in Q$ if and only if $R(x) \in Q'$;*

2. *$R$ is computable by an FPT-algorithm (with respect to $k$);*

3. *There is a computable function $g : N \to N$ such that $k'(R(x)) \leq g(k(x))$ for all $x \in \Sigma^*$.*

In addition to the FPT class, some classes of parameterized problems are defined according to their parameterized intractability level. These classes are organized in a *W-hierarchy* (FPT $\subseteq$ W[1] $\subseteq$ W[2] $\subseteq \cdots \subseteq$ W[P]), and it is conjectured that each of the containments is proper [8]. If P = NP then the hierarchy collapses [8].

We define *C-hardness* and *C-completeness* of a parameterized problem $(Q, k)$ as in classical complexity theory: $(Q, k)$ is *C*-hard under FPT-reductions if every problem in $C$ is FPT-reducible to $(Q, k)$; $(Q, k)$ is *C*-complete under FTP-reductions if $(Q, k) \in C$ and $(Q, k)$ is *C*-hard.

To cite a few examples where parameter $k$ is associated with the size of a solution, VERTEX COVER$(k)$ is FPT, CLIQUE$(k)$ is W[1]-complete, and DOMINATING SET$(k)$ is W[2]-complete (see [8]). Several other results can be found in [8].

### 3.1 The problem MIN-AND/OR$(k, r)$

By Theorem 1, MIN-AND/OR remains NP-hard even when each or-vertex has at most two out-neighbors. Let MIN-AND/OR$(k, r)$ stand for the parameterized version of MIN-AND/OR where every or-vertex of the input graph has at most $r$ out-edges with the same weight and it is asked whether there is a solution subgraph of weight at most $k$. Note that the restriction "at most $r$ out-edges with the same weight" imposed on or-vertices is in fact a far more general situation than simply restricting the out-degree of vertices to a constant. In this subsection, we show that MIN-AND/OR$(k, r)$ is in FPT for parameters $k$ and $r$.

**Theorem 3** MIN-AND/OR$(k, r)$ *is reducible to a problem kernel in time $O(m)$.*

**Proof.** The proof is based on some correct reduction rules that must be applied once in the order given below:

1. for each `and`-vertex $v_i$, if $\sum_{v_j \in O_{v_i}} \tau(v_i, v_j) > k$ then remove it;

2. for each edge $e \in E(G)$, if $\tau(e) > k$ then remove it;

3. for every vertex $v_i \neq s$, if the weight of a shortest path from $s$ to $v_i$ is greater than $k$ then remove it;

4. if some vertex becomes unreachable from $s$ then remove it;

5. for every vertex that becomes a sink, assign weight $k + 1$ to all its in-edges;

6. for each `and`-vertex such that some of its out-neighbors has been removed, assign weight $k + 1$ to all its in-edges.

Let $G'$ be the graph obtained by applying the above reduction rules. The reduction rules have modified or removed only vertices and edges that could not be part of a solution subgraph of maximum weight $k$ in $G$ and vice-versa. Thus, if $S$ is a solution subgraph of weight at most $k$ in $G'$ then $S$ is also a solution subgraph of weight at most $k$ in $G$. Note that the running time to apply the above reduction rules is $O(m)$, since $G$ is acyclic.

In $G'$ the longest shortest-path from $s$ to a sink has cost at most $k$, and each vertex has at most $kr$ out-neighbors. Thus, $G'$ will have a maximum number of vertices if: (i) all its non-sink vertices have out-degree equal to $kr$, (ii) no vertex shares a same out-neighbor with another vertex, and (iii) the cost of the shortest path from $s$ to any sink is $k$. Hence the number of vertices at distance $i$ from $s$ is at most $(kr)^i$, that is, the total number of the vertices in $G'$ is at most $O((kr)^{k+1})$.

Since (a) the reduction rules can be applied in $O(m)$ time, (b) the size of $G'$ is a function of the parameters $k$ and $r$, and (c) a solution subgraph of maximum weight $k$ in $G'$ is also a solution subgraph of maximum weight $k$ in $G$, we conclude that $G'$ is a kernel for MIN-AND/OR$(k,r)$. Hence MIN-AND/OR$(k,r)$ is reducible to a problem kernel in $O(m)$ time. □

**Corollary 4** MIN-AND/OR$(k,r)$ *is in* FPT. □

### 3.2 And/or graphs with zero-weight edges

In this subsection, we consider the family $\mathcal{Z}$ of and/or graphs where zero-weight edges are allowed. This can model practical situations in which some decisions can be taken at no cost, although in the original definition of MIN-AND/OR [18] all edges have positive weights. Let MIN-AND/OR$^0(k)$ stand for the parameterized version of Min-and/or applied to and/or graphs in $\mathcal{Z}$, and Dominating Set($c$) for the W[2]-hard parameterized problem where it is asked whether an input graph $Q$ has a dominating set of size at most $c$ (see [8]).

**Theorem 5** Dominating Set($c$) is FPT-reducible to Min-and/or$^0$($k$).

**Proof.** Given an instance $(Q, c)$ of Dominating Set($c$), we construct an instance $(G, k)$ of Min-and/or$^0$($k$) as follows: (a) create a source vertex $s$ in $G$ where $f(s) =$ and; (b) for each vertex $v_i \in V(Q)$, create three associated vertices $u_i, w_i, t_i$ where $f(u_i) =$ or, $f(w_i) =$ and, $f(t_i) =$ or; (c) for each vertex $u_i \in V(G)$, add an edge $(s, u_i)$ with $\tau(s, u_i) = 0$, and add an edge $(u_i, w_j)$ with $\tau(u_i, w_j) = 0$ if and only if $i = j$ or $(v_i, v_j) \in E(Q)$; (d) create an edge $(w_i, t_i) \in E(G)$ with $\tau(w_i, t_i) = 1$ for all $i \in \{1, \ldots, n\}$; (e) finally, set $k = c$.

If $Q$ contains a dominating set $C$ such that $|C| \leq c$ then it is possible to construct a solution subgraph $H$ of $G$ with weight at most $k$ as follows: $s$ and all of its out-neighbors belong to $V(H)$; for each vertex $u_i \in V(H)$, include in $V(H)$ an out-neighbor $w_j$ of $u_i$ if and only if $v_j \in C$; and for each vertex $w_j \in V(H)$, add an edge $(w_j, t_j)$ to $E(H)$. Since $|C| \leq c = k$ then at most $k$ edges $(w_j, t_j)$ belong to $E(H)$. Hence $H$ has weight at most $k$.

Conversely, if $G$ has a solution subgraph $H$ with weight at most $k$ then it is possible obtain a dominating set $C$ of $Q$ as follows: a vertex $v_i$ of $Q$ belongs to $C$ if and only if $w_i$ belongs to $V(H)$. Since $H$ is a solution subgraph, by definition every non-sink or-vertex has exactly one out-neighbor. Hence $H$ has at most $k$ vertices $w_i$ and $|C| \leq k$. □

Figure 3 illustrates in (a) an instance of Dominating Set and in (b) the corresponding instance of Min-and/or$^0$($k$) obtained by the construction above.

**Corollary 6** Min-and/or$^0$($k$) is W[2]-hard. □

### 3.3 The problem Min-x-y($k$)

Let Min-x-y($k$) stand for the parameterized version of Min-x-y, where it is asked whether there is a solution subgraph of weight at most $k$, and Clique($c$) for the W[1]-hard parameterized problem where it is asked whether the input graph $Q$ has a clique of size $c$ (see [8]).

**Theorem 7** Clique($c$) is FPT-reducible to Min-x-y($k$).

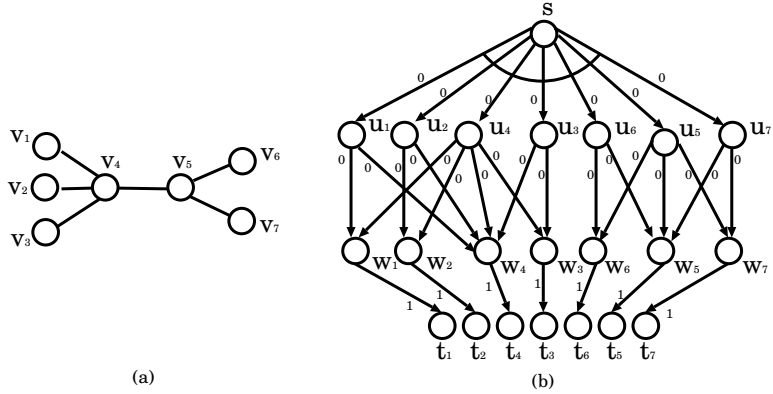

Figure 3: An instance of Dominating Set($c$) in (a), and the corresponding instance of Min-and/or$^0$($k$) in (b).

**Proof.** Given an instance $(Q, c)$ of Clique($c$), we construct an instance $(G, k)$ of Min-x-y($k$) as follows:

- create a source vertex $s$ in $G$;

- create a set $\{u_1, u_2, ..., u_n\}$ of out-neighbors of $s$, where $n = |V(Q)|$ (vertex $u_i$ of $G$ is associated with vertex $v_i$ in $Q$);

- for each vertex $u_i$, create two out-neighbors $z_i$ and $w_i$ of $u_i$;

- for each vertex $z_i$, create an edge $(z_i, w_j)$ if and only if $v_j$ and $v_i$ are neighbors in $Q$;

- for each vertex $w_i$, create an out-neighbor $t_i$ of $w_i$ ($t_i$ is a sink);

- if $v_i \in V(Q)$ has degree less than or equal to $c - 1$ then $\tau(s, u_i) = c^2 + 3c + 1$ else $\tau(s, u_i) = 1$; for all other edges in $G$ their weights are 1;

- $s$ has label $c$-$n$;

- every vertex $u_i$ has label 2-2;

- every vertex $w_i$ has label 1-1;

- every vertex $t_i$ has label 0-0;

- for each vertex $z_i$, if $d(v_i) \geq c-1$ then $z_i$ is labeled $(c-1)$-$d(v_i)$, otherwise $z_i$ is labeled $d(v_i)$-$d(v_i)$ (where $d(v_i)$ is the number of neighbors of $v_i$ in $Q$);

- set $k = c^2 + 3c$.

Figure 4 illustrates in (a) a graph $Q$, and in (b) the corresponding graph $G$.

Observe that the construction of $G$ can be done in $O(m)$ time, since $|V(G)| = 4|V(Q)| + 1$. We show that $Q$ contains a clique of size $c$ if and only if $G$ contains a solution subgraph of size less than or equal to $k$.

If $Q$ contains a set of vertices $\{v_1, v_2, ..., v_c\}$ forming a clique $C$ of size $c$, then a solution subgraph $H$ of $G$ is constructed as follows. Since $s$ is a vertex with label c-n, choose $\{u_1, u_2, ..., u_c\}$ to be the out-neighbors of $s$ in $H$. Now each vertex $u_i$ has label 2-2, and thus vertices $w_1, w_2, \ldots, w_c$ and $z_1, z_2, \ldots, z_c$ are also part of the solution subgraph $H$. This implies that vertices $t_1, t_2, \ldots, t_c$ belong to $V(H)$ as well. At this point, $H$ already contains $4c$ edges of weight 1. Since each vertex $z_i$ depends on $c-1$ out-neighbors, choose an out-neighbor $w_j$ of $z_i$ if and only if $v_j \in C$. Note that out-edges of vertices $z_1, z_2, \ldots, z_c$ add weight $c(c-1)$ to $H$. In addition, selected out-neighbors of each vertex $z_i$ were already in $H$ before their choice. Hence the weight of $H$ is $c(c-1) + 4c = c^2 + 3c = k$.

Conversely, suppose that $G$ contains an optimal solution subgraph $H$ of weight at most $k \leq c^2 + 3c$. Note that $H$ is a solution subgraph such that: (i) $s$ has $c$ out-neighbors $u_i$; (ii) each out-neighbor $u_i$ of $s$ has two out-neighbors $z_i$ and $w_i$; (iii) each one of the $c$ vertices $z_i$ has $c-1$ out-neighbors. From these observations, $H$ contains so far at least $c^2 + 2c$ edges, that is, $H$ contains at most $c$ vertices $w_i$. By construction, if $w_i \in V(H)$ then vertices $u_i$ and $z_i$ also belong to $V(H)$; but since there is no edge between $z_i$ and $w_i$, $H$ contains exactly $c$ vertices $w_i$, and $(z_i, w_j) \in E(H)$ for all $w_j \neq w_i$ belonging to $V(H)$. Let $C$ be the subset of vertices $v_i \in V(Q)$ such that $v_i \in C$ if and only if $w_i \in H$. Since $u_i, z_i, w_i$ in $G$ are associated with $v_i$ in $Q$ and out-edges of $z_i$ in $G$ represent the neighborhood of $v_i$ in $Q$, we conclude that $C$ is a clique of size $c$ in $Q$. Hence CLIQUE($c$) is FTP-reducible to MIN-X-Y($k$). $\square$

**Corollary 8** MIN-X-Y($k$) *is* W[1]*-hard.* $\square$

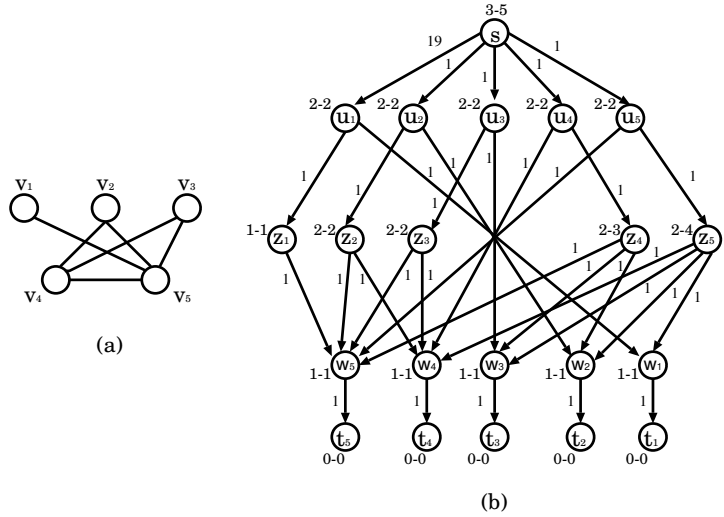

Figure 4: FPT-reduction of graph $Q$ in (a) to x-y graph $G$ in (b).

## 4 Conclusions

In this paper we have proved that Min-and/or remains NP-hard even for and/or graphs where edges have weight one, or-vertices have out-degree at most two, and vertices in with in-degree greater than one are within distance at most one of a sink; and that deciding whether there is a solution subtree with weight exactly $k$ of a given x-y tree is also NP-hard. We also have shown that Min-and/or$(k,r)$ is in FPT, Min-and/or$^0(k)$ is W[2]-hard, and Min-x-y$(k)$ is W[1]-hard.

The question of classifying the parameterized problem Min-and/or$(k)$ for and/or graphs whose edges have *positive* weights remains open.